\documentclass[12pt,prd,tightenlines,nofootinbib,showpacs,showkeys,setspace]{revtex4}
\newcommand{\be}{\begin{equation}}
\newcommand{\ee}{\end{equation}}
\newcommand{\bdis}{\begin{displaymath}}
\newcommand{\edis}{\end{displaymath}}
\newcommand{\bga}{\begin{equation}\begin{gathered}}
\newcommand{\ega}{\end{gathered}\end{equation}}
\newcommand{\mathsym}[1]{{}}
\newcommand{\unicode}[1]{{}}
\usepackage{bm}
\usepackage{graphics}
\usepackage{rotating}
\usepackage{epsfig}
\usepackage{amsmath}
\usepackage{amsfonts}
\usepackage{setspace}
\usepackage{amssymb}

\begin{document}
\title{\begin{flushright}{\rm\normalsize SSU-HEP-16/04\\[5mm]}\end{flushright}
Relativistic corrections to the pair $B_c$-meson \\production in $e^+e^-$ annihilation}
\author{\firstname{A.A.} \surname{Karyasov}}
\affiliation{Samara University, Moskovskoye Shosse 34, 443086, Samara, Russia}
\author{\firstname{A.P.} \surname{Martynenko}}
\affiliation{Samara University, Moskovskoye Shosse 34, 443086, Samara, Russia}
\author{\firstname{F.A.} \surname{Martynenko}}
\affiliation{Samara University, Moskovskoye Shosse 34, 443086, Samara, Russia}

\begin{abstract}
Relativistic corrections to the pair $B_c$-meson production in $e^+e^-$-annihilation
are calculated. We investigate a production of pair pseudoscalar, vector and pseudoscalar+vector
$B_c$-mesons in the leading order perturbative quantum chromodynamics and
relativistic quark model. Relativistic expressions of the pair production cross sections
are obtained. Their numerical evaluation shows that relativistic effects in the
production amplitudes and bound state wave functions three times reduce nonrelativistic results
at the center-of-mass energy $s=22~GeV$.
\end{abstract}

\pacs{13.66.Bc, 12.39.Ki, 12.38.Bx}

\keywords{Hadron production in $e^+e^-$ interactions, Relativistic quark model}

\maketitle

\section{Introduction}

In recent years, one of the important centers of research in the quark physics was the study of the mechanisms of heavy quarkonium production
in electron-positron annihilation, and at the Large Hadron Collider. Such studies allow us to test the Standard Model,
to clarify the values of fundamental parameters of the theory, to test theoretical models used for a description of
heavy quark bound states in quantum chromodynamics. One of the important reactions that occur in the electron-positron annihilation
is the pair production of heavy quarkoniums and diquarks \cite{pahlova,brambilla-2011}. In the study of such reactions it has been
found that significantly increases the role of the theory of the formation of bound quark states, namely, how originally
appeared two quarks and two anti-quarks at small distances then combine into mesons. It has been shown that it is
impossible to achieve a good description of experimental data without accounting effects
of relative quark motion \cite{BL1,bodwin,Chao,Qiao,BLL,EM2006,ji,jia,gong,akl}.
A system $(\bar b c)$ with open beauty and charm has a special place among the heavy quarkoniums since its decay mechanism differs significantly
from the decay mechanism of charmonium or bottomonium. At that time, as the pair charmonium and bottomium production in electron-positron
annihilation already studied both theoretically and experimentally \cite{brambilla-2011,Belle,BaBar}, the production of a pair of $B_c$
mesons is studied much less.
It has its own specific features. Their research will extend quantitative understanding of quantum chromodynamics,
the check of the bound state theory of quarks with different flavors.

This work continues the series of our works on exclusive double charmonium production in $e^+e^-$ annihilation
to the case of $B_c$-meson production. The mechanism of pair $B_c$ meson production in electron-positron annihilation
is more simple in comparison with other reaction of proton-proton interaction.
Our approach to the calculation of the observed cross sections for pair
production of mesons is based on methods of relativistic quark model (RQM) and perturbative quantum chromodynamics
\cite{EM2006,EFGM2009,EM2010,apm2005,rqm5,rqm11,mt,mt1}. This approach allows a systematic account of relativistic effects
as in the construction of relativistic amplitudes of pair production of mesons, relativistic production cross sections,
and in the description of bound states of quarks themselves through the use of the corresponding quark interaction potential.
In general, this approach creates a microscopic picture of the interaction of quarks at different stages of meson production.
It also allows you to perform a self-consistent calculation of numerous parameters that define the cross section
for meson production, from relativistic parameters determining the movement of heavy quarks, to the
masses themselves of quark bound states. We can say that one of the main problems of the theory of strong
interactions - a significant increase in the phenomenological parameters of various types, is solved in
this case within the model. The aim of our work is to extend previously used methods for pair charmonium production
to the case of quarks of different flavors. The question naturally arises of whether there can be the pair $B_c$ meson production
processes in electron-positron annihilation quite probable that they can be observed in the experiment. In our work
we try to show that the process of pair production of $B_c$ mesons have a clear experimental
prospect.

\section{General formalism}

\begin{figure}[t!]
\centering
\includegraphics[width=5.0 cm]{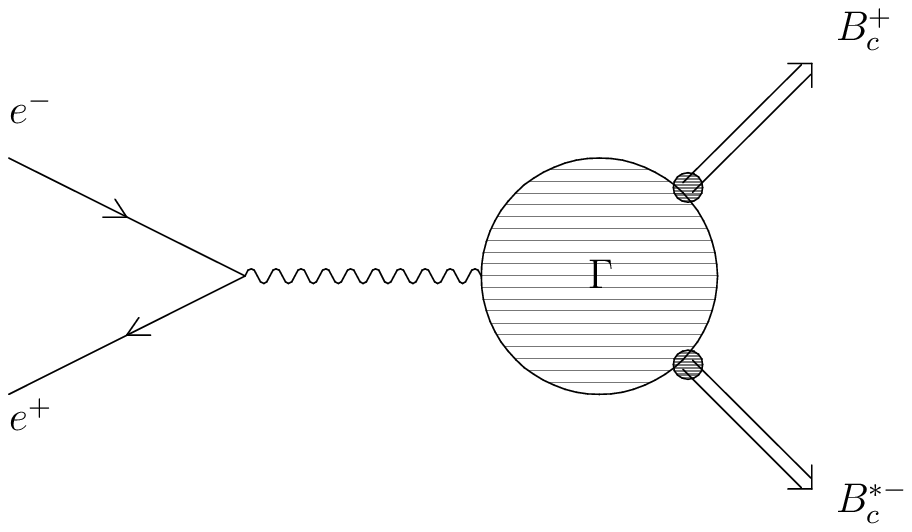}\hspace*{0.4cm}
\includegraphics[width=11.0 cm]{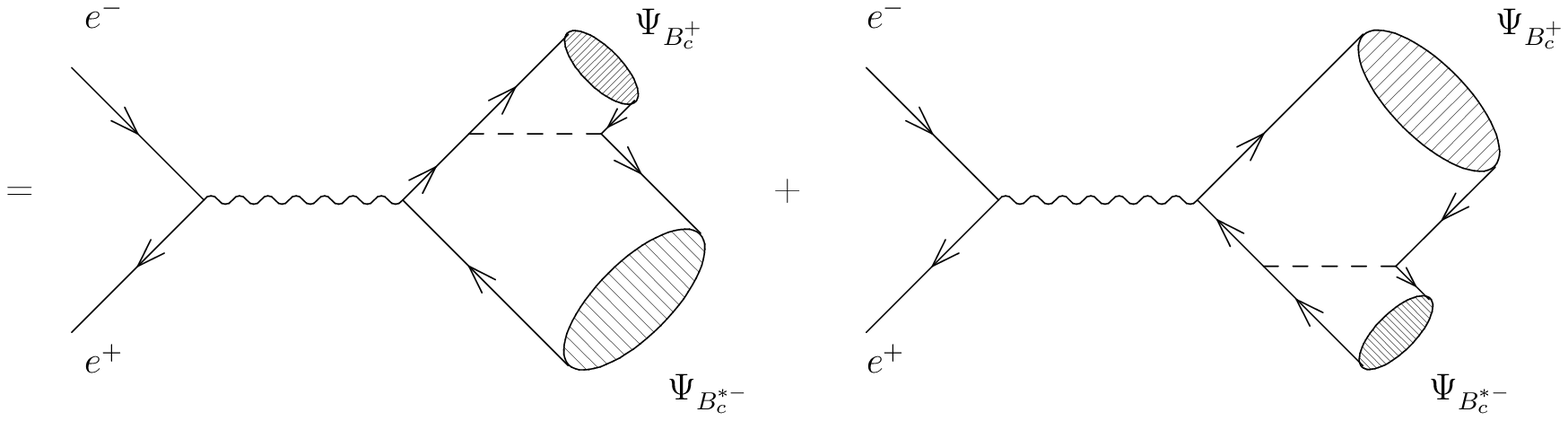}
\caption{The pair $B_c$-meson production amplitudes in $e^+e^-$ annihilation. $B^+_{c}$ and $B^{\ast-}_{c}$
denote the $B_c$-meson states with spin 0 and 1. Wavy line shows
the virtual photon and dashed line corresponds to the gluon. $\Gamma$ is the production vertex function.}
\label{fig:fig1}
\end{figure}

Two production amplitudes of the $B_c$ meson pair in leading order of the QCD coupling constant $\alpha_s$ are presented
in Fig.~\ref{fig:fig1}. Two other amplitudes can be obtained by corresponding permutations.
There are two stages of $B_c$ meson production process \cite{braaten,kramer}. At the
first process step, which is described by perturbative QCD, the virtual photon $\gamma^\ast$ and then virtual gluon
$g^\ast$ produce two heavy quarks $(bc)$ and two heavy antiquarks $(\bar b\bar c)$ with the following four-momenta:
\begin{equation}
\label{eq:pq}
p_1=\eta_{1}P+p,~p_2=\eta_{2}P-p,~(p\cdot P)=0,~\eta_{i}=\frac{M_{B_{\bar bc}}^2\pm m_1^2\mp m_2^2}{2M_{B_{\bar bc}}^2},
\end{equation}
\begin{displaymath}
q_1=\rho_{1}Q+q,~q_2=\rho_{2}Q-q,~(q\cdot Q)=0,~\rho_{i}=\frac{M_{B_{b\bar c}}^2\pm m_1^2\mp m_2^2}{2M_{B_{b\bar c}}^2}
\end{displaymath}
where $M_{B_{\bar bc}}$ is the mass of pseudoscalar $B_c^+$ or vector $B_c^{\ast +}$ meson consisting of $\bar b$-antiquark 
and $c$-quark.
$P(Q)$ are the total four-momenta of mesons $B_c^+$ and $B_c^{\ast -}$, relative quark four-momenta $p=L_P(0,{\bf p})$ and
$q=L_P(0,{\bf q})$ are obtained from the rest frame four-momenta $(0,{\bf p})$ and $(0,{\bf q})$ by the
Lorentz transformation to the system moving with the momenta $P$ and $Q$.
The index $i=1,2$ corresponds to plus and minus signs in \eqref{eq:pq}.
Heavy quarks $c$, $b$ and antiquarks $\bar c$, $ \bar b$ in the intermediate state are outside the mass shell:
$p_{1,2}^2=\eta_{i}^2P^2-{\bf p}^2=\eta_{i}^2M_{B_{\bar bc}}^2-{\bf p}^2\not= m_{1,2}^2$,
so that $p_1^2-m_1^2=p_2^2-m_2^2$. At the second nonperturbative step of the production process, quark-antiquark pairs form
double heavy mesons of definite spin.

Let consider the production amplitude of pseudoscalar and vector $B_c$ mesons. Initially it can be written as a
convolution of perturbative production amplitude of free quarks and antiquarks and the quasipotential wave
functions. Using then the transformation law of the bound state wave functions from the rest frame to the
moving one with four-momenta $P$ and $Q$ we can present the meson production amplitude in the form \cite{rqm5,EM2006,EM2010}:
\begin{equation}
\label{eq:amp}
{\cal M}(p_-,p_+,P,Q)=\frac{8\pi^2\alpha}{3s^2}\sqrt{M_{B_{\bar bc}}M_{B_{b\bar c}}}\bar v(p_+)\gamma^\beta u(p_-)
\int\frac{d{\bf p}}{(2\pi)^3}
\int\frac{d{\bf q}}{(2\pi)^3}
\times
\end{equation}
\begin{displaymath}
\times Sp\left\{\Psi^{\cal P}_{B_{\bar bc}}(p,P)\Gamma_1^{\beta\nu}(p,q,P,Q)\Psi^{\cal V}_{B_{b\bar c}}(q,Q)\gamma_\nu+
\Psi^{\cal P}_{B_{\bar bc}}(-p,P)\Gamma_2^{\beta\nu}(p,q,P,Q)\Psi^{\cal V}_{B_{b\bar c }}(-q,Q)
\gamma_\nu\right\},
\end{displaymath}
where $s$ is the center-of-mass energy,
a superscript ${\cal P}$ indicates a pseudoscalar $B_c$ meson, a superscript ${\cal V}$ indicates a vector
$B_c$ meson, $\alpha$ is the fine structure constant. $\Gamma_{1,2}$ are the vertex functions defined below.
The permutation of subscripts $b$ and $c$ in the wave functions indicates corresponding permutation
in the projection operators (see below Eqs.\eqref{eq:amp1}-\eqref{eq:amp2}.
The method for producing the amplitudes in the form \eqref{eq:amp} is described in our previous studies
\cite{apm2005,EM2006,EM2010}.
The transition of free quark-antiquark pair to meson bound states is described in our approach by specific wave functions.
Relativistic wave functions of pseudoscalar and vector $B_c$ mesons accounting for the transformation from the
rest frame to the moving one with four momenta $P$, and $Q$ are
\begin{eqnarray}
\label{eq:amp1}
\Psi^{\cal P}_{B_{\bar bc}}(p,P)&=&\frac{\Psi^0_{B_{\bar bc}}({\bf p})}{
\sqrt{\frac{\epsilon_1(p)}{m_1}\frac{(\epsilon_1(p)+m_1)}{2m_1}
\frac{\epsilon_2(p)}{m_2}\frac{(\epsilon_2(p)+m_2)}{2m_2}}}
\left[\frac{\hat v_1-1}{2}+\hat
v_1\frac{{\bf p}^2}{2m_2(\epsilon_2(p)+ m_2)}-\frac{\hat{p}}{2m_2}\right]\cr
&&\times\gamma_5(1+\hat v_1) \left[\frac{\hat
v_1+1}{2}+\hat v_1\frac{{\bf p}^2}{2m_1(\epsilon_1(p)+
m_1)}+\frac{\hat{p}}{2m_1}\right],
\end{eqnarray}
\begin{eqnarray}
\label{eq:amp2}
\Psi^{\cal V}_{B^\ast_{b\bar c}}(q,Q)&=&\frac{\Psi^0_{B^\ast_{b\bar c}}({\bf q})}
{\sqrt{\frac{\epsilon_1(q)}{m_1}\frac{(\epsilon_1(q)+m_1)}{2m_1}
\frac{\epsilon_2(q)}{m_2}\frac{(\epsilon_2(q)+m_2)}{2m_2}}}
\left[\frac{\hat v_2-1}{2}+\hat v_2\frac{{\bf q}^2}{2m_1(\epsilon_1(q)+
m_1)}+\frac{\hat{q}}{2m_1}\right]\cr &&\times\hat{\varepsilon}_{\cal
V}(Q,S_z)(1+\hat v_2) \left[\frac{\hat v_2+1}{2}+\hat
v_2\frac{{\bf q}^2}{2m_2(\epsilon_2(q)+ m_2)}-\frac{\hat{q}}{2m_2}\right],
\end{eqnarray}
where the symbol hat denotes convolution of four-vector with the Dirac gamma matrices,
$v_1=P/M_{B_{\bar bc}}$, $v_2=Q/M_{B_{b\bar c}}$;
$\varepsilon_{\cal V}(Q,S_z)$ is the polarization vector of the $B^{\ast-}_c(1^-)$ meson,
relativistic quark energies $\epsilon_{1,2}(p)=\sqrt{p^2+m_{1,2}^2}$ and $m_{1,2}$
are the masses of $c$ and $b$ quarks. Relativistic functions~\eqref{eq:amp1}-\eqref{eq:amp2}
and the vertex production functions $\Gamma_{1,2}$
do not contain the $\delta ({\bf p}^2-\eta_{i}^2M_{B_{\bar bc}}^2+m_{1,2}^2)$ which corresponds to the transition on the mass shell.
In~\eqref{eq:amp1} and \eqref{eq:amp2} we have complicated factor including the bound state wave function in the rest frame.
Therefore instead of the substitutions $M_{B_{\bar bc}}=\epsilon_1({\bf p})+\epsilon_2({\bf p})$ and
$M_{B^\ast_{b\bar c}}=\epsilon_1({\bf q})+\epsilon_2({\bf q})$ in the production amplitude we carry out the
integration over the quark relative momenta ${\bf p}$ and ${\bf q}$.
Color part of the meson wave function in the amplitude~\eqref{eq:amp} is taken as $\delta_{ij}/\sqrt{3}$
(color indexes $i, j, k=1, 2, 3$).
Relativistic wave functions in~\eqref{eq:amp1} and \eqref{eq:amp2} are equal to the product of wave functions in the rest frame
$\Psi^0_{B_{\bar bc}}({\bf p})$ and spin projection operators that are
accurate at all orders in $|{\bf p}|/m$ \cite{rqm5,EM2006}. An expression of spin projector in different
form for $(c\bar c)$ system was obtained in \cite{Bodwin2002} where spin projectors are
written in terms of heavy quark momenta $p_{1,2}$ lying on the mass shell.
Our derivation of relations~\eqref{eq:amp1} and \eqref{eq:amp2} accounts for the transformation
law of the bound state wave functions from the rest frame to the
moving one with four momenta $P$ and $Q$. This transformation law was discussed in the Bethe-Salpeter
approach in \cite{BP} and in quasipotential method in \cite{F1973}.

We have omitted here intermediate expressions, leading to the equations~\eqref{eq:amp}-\eqref{eq:amp2}
because they were discussed in detail in our previous papers \cite{EM2006,EFGM2009}.
In the Bethe-Salpeter approach the initial production amplitude has a form of convolution of the truncated
amplitude with two Bethe-Salpeter (BS) $B_c$ meson wave functions.
The presence of the $\delta (p\cdot P)$ function in this case
allows us to make the integration over relative energy $p^0$. In the rest frame of a bound state the condition
$p^0=0$ allows to eliminate the relative energy
from the BS wave function. The BS wave function satisfies a two-body bound state equation
which is very complicated and has no known solution. A way to deal with this problem
is to find a soluble lowest-order equation containing main physical properties
of the exact equation and develop a perturbation theory. For this purpose we continue
to work in three-dimensional quasipotential approach. In this framework the double
$B_c$ meson production amplitude~\eqref{eq:amp} can be written initially as a product of the production
vertex function $\Gamma_{1,2}$ projected onto the positive energy states by means of the Dirac
bispinors (free quark wave functions) and a bound state quasipotential wave functions
describing $B_c$ mesons in the reference frames moving with four momenta $P,Q$.
Further transformations include the known transformation law of the bound state wave
functions to the rest frame \cite{EM2006,EFGM2009}. The physical
interpretation of the double $B_c$ production amplitude is the following:
we have a complicated transition of two heavy quark and antiquark pairs
which are produced in $e^+e^-$-annihilation outside the mass shell and their
subsequent evolution firstly on the mass shell (free Dirac bispinors) and then to the
meson states. In the spin projectors we have
${\bf p}^2\not=\eta_{i}^2M^2-m_{1,2}^2$ just the same as in the vertex production functions
$\Gamma_{1,2}$.
We can consider \eqref{eq:amp1}-\eqref{eq:amp2} as a transition form factors for
heavy quark-antiquark pair from free state to bound state. When transforming the amplitude ${\cal M}$
we introduce the projection operators $\hat\Pi^{\cal P,V}$ onto the states of $(Q_1\bar Q_2)$ in the $B_c$ meson
with total spin 0 and 1 as follows:
\begin{equation}
\label{eq:uu}
\hat\Pi^{\cal P,V}=[v_2(0)\bar u_1(0)]_{S=0,1}=\gamma_5(\hat\varepsilon^\ast)\frac{1+\gamma^0}{2\sqrt{2}}.
\end{equation}

At leading order in $\alpha_s$ the vertex functions $\Gamma_{1,2}^{\beta\nu}(p,P;q,Q)$ can be written as
($\Gamma_2^{\beta\nu}(p,P;q,Q)$ can be obtained from $\Gamma_1^{\beta\nu}(p,P;q,Q)$ by means of the replacement
$p_1\leftrightarrow p_2$, $q_1\leftrightarrow q_2$, $\alpha_b\to\alpha_c$, $Q_c\to Q_b$)
\begin{equation}
\label{eq:g1}
\Gamma_1^{\beta\nu}(p,P;q,Q)= Q_c\alpha_b\left[\gamma_\mu\frac{(\hat l-\hat
q_1+m_1)}{(l-q_1)^2-m_1^2+i\epsilon} \gamma_\beta +
\gamma_\beta\frac{(\hat p_1-\hat l+m_1)}{(p_1-l)^2-m_1^2+i\epsilon}
\gamma_\mu\right]D^{\mu\nu}(k_2),
\end{equation}
\begin{equation}
\label{eq:g2}
\Gamma_2^{\beta\nu}(p,P;q,Q)=Q_b\alpha_c\left[\gamma_\mu\frac{(\hat l-\hat q_2+m_2)}{(l-q_2)^2-m_2^2+i\epsilon} \gamma_\beta+
\gamma_\beta\frac{(\hat p_2-\hat l+m_2)}{(p_2-l)^2-m_2^2+i\epsilon}
\gamma_\mu\right]D^{\mu\nu}(k_1),
\end{equation}
where the square of total four-momentum $l^2$ can be expressed in terms of four momenta of the electron and positron
$p_-$, $p_+$ and four-momenta of $B_c$ mesons: $l^2=s^2=(P+Q)^2=(p_-+p_+)^2$,
the gluon four-momenta are $k_1=p_1+q_1$, $k_2=p_2+q_2$, $\alpha_{c,b}=\alpha_s\left(\frac{m_{1,2}^2}{M^2}s^2\right)$.
Relative momenta $p$, $q$ of heavy quarks enter in the gluon propagators $D_{\mu\nu}(k_{1,2})$
and quark propagators as well as in relativistic wave functions~\eqref{eq:amp1} and \eqref{eq:amp2}.
Accounting for the small ratio of relative quark momenta $p$ and $q$ to the energy $s$, we use an expansion of
inverse denominators of quark and gluon propagators as follows:
\begin{equation}
\label{eq:pr1}
\frac{1}{(l-q_{1,2})^2-m_{1,2}^2}=\frac{1}{r_{2,1}s^2}\left[Z_{1,2}-\frac{q^2-2qP}{r_{2,1}s^2}
+\cdots\right],
\end{equation}
\begin{equation}
\label{eq:pr2}
\frac{1}{(l-p_{1,2})^2-m_{1,2}^2}=\frac{1}{r_{2,1}s^2}\left[Z_{3,4}-\frac{p^2-2pQ}{r_{2,1}s^2}
+\cdots\right],
\end{equation}
\begin{equation}
\label{eq:pr3}
Z_1=\frac{r_2s^2}{\rho_1^2M_P^2+\rho_2^2s^2+\rho_1\rho_2(s^2+M_P^2-M_V^2)-m_1^2},
\end{equation}
\begin{equation}
\label{eq:pr4}
Z_3=\frac{r_1s^2}{\rho_2^2M_V^2+\rho_1^2s^2+\rho_1\rho_2(s^2+M_P^2-M_V^2)-m_2^2},
\end{equation}
\begin{equation}
\label{eq:pr5}
\frac{1}{k_{1,2}^2}=\frac{1}{r_{2,1}^2s^2}\left[Y_{1,2}\pm \frac{2r_{2,1}(pQ+qP)\mp p^2 \mp q^2 \mp 2pq}{r_{2,1}^2s^2}+\cdots\right],
\end{equation}
\begin{displaymath}
Y_{1,2}=\frac{r^2_{2,1}s^2}{\eta^2_{2,1}M^2_V+\rho^2_{2,1}M^2_P+\eta_{2,1}\rho_{2,1}(s^2-M^2_V-M^2_P)},
\end{displaymath}
where $M=m_1+m_2$, $r_{1,2}=m_{1,2}/M$. In a purely non-relativistic approximation factors $Z_i$ and $Y_i$ are equal to 1.
They accumulate the bound state effects.
The expressions for $Z_2$ and $Z_4$ can be obtained from $Z_1$ and $Z_3$ after replacement $r_2\to r_1$, $\rho_{1,2}\leftrightarrow \eta_{2,1}$, $m_1\to m_2$.
These expansions contain also the $B_c$ meson masses which can be calculated in quark model or taken from experiment.
Using expansions \eqref{eq:pr1}-\eqref{eq:pr5} and wave functions \eqref{eq:amp1}-\eqref{eq:amp2}
in the amplitude~\eqref{eq:amp} we hold the second-order correction for small ratios
$|{\bf p}|/m_{1,2}$, $|{\bf q}|/m_{1,2}$, $|{\bf p}|/s$, $|{\bf q}|/s$ relative to the leading order result.
As we take relativistic factors in the denominator of the amplitudes \eqref{eq:amp1} and \eqref{eq:amp2} unchanged,
the momentum integrals are convergent. Calculating the trace in obtained expression
in the package FORM \cite{FORM}, we find relativistic amplitudes of the $B_c$ meson pairs production
in the form:
\begin{equation}
\label{eq:amp11}
{\cal M}_{PP}=\frac{256\pi^2\alpha MM_{\cal P}}{3s^6}(v_1-v_2)^\beta\bar v(p_+)
\gamma_\beta u(p_-)\Psi^0_{B_{\bar b c}}(0)\Psi^0_{B_{b\bar c}}(0)\times
\end{equation}
\begin{displaymath}
\left[\frac{Q_c\alpha_s(\frac{m_2^2}{M^2}s^2)}{r_2^3}F_{1}^{\cal P}-
\frac{Q_b\alpha_s(\frac{m_1^2}{M^2}s^2)}{r_1^3}F_{2}^{\cal P}
\right]
\end{displaymath}
\begin{equation}
\label{eq:amp22}
{\cal M}_{PV}=\frac{256\pi^2\alpha M\sqrt{M_{\cal P}M_{\cal V}}}{3s^6}
\bar v(p_+)\gamma_\beta u(p_-)\varepsilon_{\beta\alpha\sigma\lambda}
\varepsilon_{\cal V}^\alpha v_1^\sigma v_2^\lambda
\Psi^0_{B^\ast_{\bar bc}}(0)\Psi^0_{B_{b\bar c}}(0)\times
\end{equation}
\begin{displaymath}
\left[\frac{Q_c\alpha_s(\frac{m_2^2}{M^2}s^2)}{r_2^3}F_{1}^{\cal P,V}+
\frac{Q_b\alpha_s(\frac{m_1^2}{M^2}s^2)}{r_1^3}F_{2}^{\cal P,V}\right],
\end{displaymath}
\begin{equation}
\label{eq:amp33}
{\cal M}_{VV}=\frac{256\pi^2\alpha MM_{\cal V}}{3s^6}
\bar v(p_+)\gamma_\beta u(p_-)\Psi^0_{B^\ast_{\bar bc}}(0)\Psi^0_{B^\ast_{b\bar c}}(0)\times
\end{equation}
\begin{displaymath}
\left[\frac{Q_c\alpha_s(\frac{m_2^2}{M^2}s^2)}{r_2^3}F_{1}^{\cal V,\beta}-
\frac{Q_b\alpha_s(\frac{m_1^2}{M^2}s^2)}{r_1^3}F_{2}^{\cal V,\beta}\right],
\end{displaymath}
where $\varepsilon_{{\cal V}}$ is the polarization vector of spin 1 $B_c$ meson.
The functions $F_{i}^{\cal P}$, $F_{i}^{\cal P,V}$, $F_{i}^{\cal V,\beta}$ entering in
\eqref{eq:amp11}-\eqref{eq:amp33}
can be written as series in specific relativistic factors $C_{ij}=[(m_1-\epsilon_1(p))/(m_1+\epsilon_1(p))]^i
[(m_2-\epsilon_2(q))/(m_2+\epsilon_2(q))]^j$ with $i+j\leq 2$.
To preserve a symmetry in quark masses we make following substitution in some expansion
terms: ${\bf p}^2/4m_1m_2\approx\sqrt{(\epsilon_1-m_1)(\epsilon_2-m_2)/(\epsilon_1+m_1)(\epsilon_2+m_2)}
[1+(\epsilon_1-m_1)/(\epsilon_1+m_1)+(\epsilon_2-m_2)/(\epsilon_2+m_2)]$.
Exact analytical expressions for these functions are presented in Appendix A.
To calculate the differential cross sections for the pair production, we introduce the angle
$\theta$ between the electron momentum ${\bf p}_e$ and momentum ${\bf P}$ of $B_c$ meson. Then we can obtain
the differential cross section $d\sigma/d\cos\theta$ and total cross section $\sigma$ as a
function of center-of-mass energy $s$, masses of quarks and $B_c$ mesons and relativistic
parameters presented below.
The differential cross sections for pair $B_c$ meson production can be written in the following form:
\begin{equation}
\label{eq:sech1}
\frac{d\sigma_{PP}}{d\cos\theta}=\frac{1024\pi^3\alpha^2}{9s^{10}}M_P^{2}
|\Psi^0_{B_{\bar bc}}(0)|^2|\Psi_{{B}_{b\bar c}}(0)|^2\left(1-
\frac{4M^2_P}{s^2}\right)^{3/2}\left(1-\cos^2\theta\right)\times
\end{equation}
\begin{displaymath}
\left[\frac{Q_c\alpha_s(\frac{m_2^2}{M^2}s^2)}{r_2^3}F_{1}^{\cal P}-
\frac{Q_b\alpha_s(\frac{m_1^2}{M^2}s^2)}{r_1^3}F_{2}^{\cal P}\right]^2,
\end{displaymath}
\begin{equation}
\label{eq:sech2}
\frac{d\sigma_{PV}}{d\cos\theta}=\frac{256\pi^3\alpha^2}{9s^8}\frac{M_PM_V}{M^2}
\left[\left(1-\frac{(M_P+M_V)^2}{s^2}\right)\left(1-\frac{(M_P-M_V)^2}{s^2}\right)\right]^{3/2}\times
\end{equation}
\begin{displaymath}
|\Psi^0_{B^\ast_{\bar bc}}(0)|^2|\Psi_{B_{b\bar c}}(0)|^2
\left[\frac{Q_c\alpha_s\left(\frac{m_2^2}{M^2}s^2\right)}{r_2^3}F_{1}^{\cal P,V}+
\frac{Q_b\alpha_s\left(\frac{m_1^2}{M^2}s^2\right)}{r_1^3}F_{2}^{\cal P,V}\right]^2
\left(2-\sin^2\theta\right),
\end{displaymath}
\begin{equation}
\label{eq:sech3}
\frac{d\sigma_{VV}}{d\cos\theta}=\frac{256\pi^3\alpha^2}{9s^{10}}M_V^2
|\Psi^0_{B^\ast_{\bar bc}}(0)|^2|\Psi^0_{B^\ast_{b\bar c}}(0)|^2
\left(1-\frac{4M^2_V}{s^2}\right)^{3/2}\left(F_A-F_B\cdot\cos^2\theta\right),
\end{equation}
\begin{displaymath}
F_A=F^2_{1,V}(12-4\eta+\eta^2)+F_{1,V}F_{2,V}(8\eta-6\eta^2+\eta^3)+F_{1,V}F_{3,V}(4\eta-2\eta^2)+
\end{displaymath}
\begin{displaymath}
+F^2_{2,V}(4\eta^2-2\eta^3+\frac{1}{4}\eta^4)+F_{2,V}F_{3,V}(4\eta^2-\eta^3)+F^2_{3,V}(2\eta+\eta^2),
\end{displaymath}
\begin{displaymath}
F_B=F^2_{1,V}(12-4\eta+\eta^2)+F_{1,V}F_{2,V}(8\eta-6\eta^2+\eta^3)+F_{1,V}F_{3,V}(4\eta-2\eta^2)+
\end{displaymath}
\begin{displaymath}
+F^2_{2,V}(4\eta^2-2\eta^3+\frac{1}{4}\eta^4)+F_{2,V}F_{3,V}(4\eta^2-\eta^3)+F^2_{3,V}(-2\eta+\eta^2),
\end{displaymath}
\begin{displaymath}
F_{i,V}=\frac{Q_c\alpha_s(\frac{m_2^2}{M^2}s^2)}{r_2^3}F_{1i}^{\cal V}-\frac{Q_b\alpha_s(\frac{m_1^2}{M^2}s^2)}{r_1^3}F_{2i}^{\cal V},
\end{displaymath}
where $\eta=s^2/M^2_V$, the values of wave function at the origin are equal
\begin{equation}
\Psi^0_{B_{\bar bc}}(0)=\int \sqrt{\frac{(\epsilon_1(p)+m_1)(\epsilon_2(p)+m_2)}
{2\epsilon_1(p)\cdot 2\epsilon_2(p)}}\Psi^0_{B_{\bar bc}}({\bf p})\frac{d{\bf p}}{(2\pi)^3}.
\end{equation}

The differential cross sections \eqref{eq:sech1}-\eqref{eq:sech3} in this form are very close to nonrelativistic results
obtained in \cite{kiselev}.
The functions $F_{i}^{\cal P}$, $F_{i}^{\cal P,V}$ and $F_{ij}^{\cal V}$ are obtained as series in $|{\bf p}|/m_{1,2}$ and
$|{\bf q}|/m_{1,2}$ up to corrections of second order.
The functions $F_{i}^{\cal P}$, $F_{i}^{\cal P,V}$ and $F_{ij}^{\cal V}$ (see Appendix A) contain
relativistic parameters $\omega^{P,V}_{nk}$ which can be expressed in terms of momentum integrals $I_{nk}$ and
calculated in the quark model:
\begin{equation}
\label{eq:intnk}
I_{nk}^{P,V}=\int_0^\infty q^2R^{P,V}_{D_{bc}}(q)\sqrt{\frac{(\epsilon_1(q)+m_1)(\epsilon_2(q)+m_2)}
{2\epsilon_1(q)\cdot 2\epsilon_2(q)}}
\left(\frac{m_1-\epsilon_1(q)}{m_1+\epsilon_1(q)}\right)^n
\left(\frac{m_2-\epsilon_2(q)}{m_2+\epsilon_2(q)}\right)^k dq,
\end{equation}
\begin{equation}
\label{eq:parameter}
\omega^{P,V}_{10}=\frac{I^{P,V}_{10}}{I^{P,V}_{00}},~~~\omega^{P,V}_{01}=\frac{I^{P,V}_{01}}{I^{P,V}_{00}},
~~~\omega^{P,V}_{\frac{1}{2}\frac{1}{2}}=\frac{I^{P,V}_{\frac{1}{2}\frac{1}{2}}}{I^{P,V}_{00}},~~~
\omega^{P,V}_{20}=\frac{I^{P,V}_{20}}{I^{P,V}_{00}},
\end{equation}
\begin{displaymath}
\omega^{P,V}_{02}=\frac{I^{P,V}_{02}}{I^{P,V}_{00}},~~~\omega^{P,V}_{11}=\frac{I^{P,V}_{11}}{I^{P,V}_{00}},~~~
\omega^{P,V}_{\frac{3}{2}\frac{1}{2}}=\frac{I^{P,V}_{\frac{3}{2}\frac{1}{2}}}{I^{P,V}_{00}},~~~
\omega^{P,V}_{\frac{1}{2}\frac{3}{2}}=\frac{I^{P,V}_{\frac{1}{2}\frac{3}{2}}}{I^{P,V}_{00}}.
\end{displaymath}

There are two sources of relativistic corrections to the production amplitude~\eqref{eq:amp} and the
cross sections \eqref{eq:sech1}, \eqref{eq:sech2} and \eqref{eq:sech3} connected with relative motion of
heavy quarks. Firstly, there are various factors containing relative momenta of heavy quarks $p$ and $q$.
In final expressions they are defined by specific relativistic parameters $\omega^{P,V}_{nk}$.
Corresponding momentum integrals \eqref{eq:intnk} are convergent. They are calculated
numerically, using the bound state wave functions obtained by numerical solution of the Schr\"odinger
equation. Despite the convergence of integrals for their calculation we introduce the cutoff parameter
$\Lambda\approx m_c$ in momentum integrals $I_{nk}$ \eqref{eq:intnk} at high relative momenta $q$ since the wave
functions are not known exactly at relativistic momenta.

Secondly, there are relativistic corrections to the bound state wave functions of pseudoscalar and vector $B_c$ mesons.
The exact form of the bound state wave functions $\Psi^0_{B_{\bar bc}}({\bf q})$ and $\Psi^0_{B^\ast_{\bar bc}}({\bf q})$
is important to take into account this source of relativistic corrections. As it follows from \eqref{eq:sech1},
\eqref{eq:sech2} and \eqref{eq:sech3} in nonrelativistic approximation the pair $B_c$ meson production cross sections
contain fourth power of nonrelativistic wave function at the origin. The value of the cross sections is very sensitive to
small changes of $\Psi^0_{B_{\bar bc}}$. In nonrelativistic QCD there exists corresponding problem of determining the magnitude of the
color-singlet matrix elements \cite{BBL}. To account for the relativistic corrections in the calculation of the wave functions
we believe that the dynamics of heavy quarks is determined by the QCD generalization of
the standard Breit Hamiltonian in the center-of-mass reference frame \cite{repko1,pot1,pot3,capstick}:
\begin{equation}
\label{eq:breit}
H=H_0+\Delta U_1+\Delta U_2,~~~H_0=\sqrt{{\bf
p}^2+m_1^2}+\sqrt{{\bf p}^2+m_2^2}-\frac{4\tilde\alpha_s}{3r}+(Ar+B),
\end{equation}
\begin{equation}
\label{eq:breit1}
\Delta U_1(r)=-\frac{\alpha_s^2}{3\pi r}\left[2\beta_0\ln(\mu
r)+a_1+2\gamma_E\beta_0
\right],~~a_1=\frac{31}{3}-\frac{10}{9}n_f,~~\beta_0=11-\frac{2}{3}n_f,
\end{equation}
\begin{equation}
\label{eq:breit2}
\Delta U_2(r)=-\frac{2\alpha_s}{3m_1m_2r}\left[{\bf p}^2+\frac{{\bf
r}({\bf r}{\bf p}){\bf p}}{r^2}\right]+\frac{2\pi
\alpha_s}{3}\left(\frac{1}{m_1^2}+\frac{1}{m_2^2}\right)\delta({\bf r})+
\frac{4\alpha_s}{3r^3}\left(\frac{1}{2m_1^2}+\frac{1}{m_1m_2}\right)({\bf S}_1{\bf L})+
\end{equation}
\begin{displaymath}
+\frac{4\alpha_s}{3r^3}\left(\frac{1}{2m_2^2}+\frac{1}{m_1m_2}\right)({\bf S}_2{\bf L})
+\frac{32\pi\alpha_s}{9m_1m_2}({\bf S}_1{\bf S}_2)\delta({\bf r})+
\frac{4\alpha_s}{m_1m_2r^3}\left[\frac{({\bf S}_1{\bf r})({\bf S}_2{\bf r})}{r^2}-
\frac{1}{3}({\bf S}_1{\bf S}_2)\right]-
\end{displaymath}
\begin{displaymath}
-\frac{\alpha_s^2(m_1+m_2)}{m_1m_2r^2}\left[1-\frac{4m_1m_2}{9(m_1+m_2)^2}\right],
\end{displaymath}
where ${\bf L}=[{\bf r}\times{\bf p}]$, ${\bf S}_1$, ${\bf S}_2$ are spins of heavy quarks,
$n_f$ is the number of flavors, $\gamma_E\approx 0.577216$ is
the Euler constant. To improve an agreement of theoretical hyperfine splittings in $(\bar bc)$ mesons
with experimental data and other calculations in quark models we add to the standard Breit potential \eqref{eq:breit2}
the spin confining potential obtained in \cite{repko1,repko2}:
\begin{equation}
\Delta V^{hfs}_{conf}(r)=
f_V\frac{A}{8r}\left\{\frac{1}{m_1^2}+\frac{1}{m_2^2}+\frac{16}{3m_1m_2}({\bf S}_1{\bf S}_2)+
\frac{4}{3m_1m_2}\left[3({\bf S}_1 {\bf r}) ({\bf S}_2 {\bf r})-({\bf S}_1 {\bf S}_2)\right]\right\},
\end{equation}
where we take the parameter $f_V=0.9$. For the dependence of the
QCD coupling constant $\tilde\alpha_s(\mu^2)$ on the renormalization point
$\mu^2$ in the pure Coulomb term in~\eqref{eq:breit} we use the three-loop result \cite{kniehl1997}
\begin{equation}
\tilde\alpha_s(\mu^2)=\frac{4\pi}{\beta_0L}-\frac{16\pi b_1\ln L}{(\beta_0 L)^2}+\frac{64\pi}{(\beta_0L)^3}
\left[b_1^2(\ln^2 L-\ln L-1)+b_2\right], \quad L=\ln(\mu^2/\Lambda^2).
\end{equation}
In other terms of the Hamiltonians~\eqref{eq:breit1} and \eqref{eq:breit2} we use
the leading order approximation for $\alpha_s$. The typical momentum transfer scale in a
quarkonium is of order of double reduced mass, so we set the renormalization scale $\mu=2m_1m_2/(m_1+m_2)$ and
$\Lambda=0.168$ GeV, which gives $\alpha_s=0.265$ for $(\bar bc)$ meson.
The coefficients $b_i$ are written explicitly in \cite{kniehl1997}.
The parameters of the linear potential $A=0.18$ GeV$^2$ and $B=-0.16$ GeV have established values in quark models.

\begin{table}[h]
\caption{Numerical values of relativistic parameters~\eqref{eq:parameter}
in pair $B_c$ meson production cross sections~\eqref{eq:sech1}, \eqref{eq:sech2}, \eqref{eq:sech3}.}
\bigskip
\label{tb1}
\begin{ruledtabular}
\begin{tabular}{|c|c|c|c|c|c|c|c|c|c|c|c|}
$B_c$ &$n^{2S+1}L_J$ &$M_{B_{\bar b c}}$, &$\Psi^0_{B_{\bar b c}}(0)$, & $\omega^{P,V}_{10}$ &$\omega^{P,V}_{01}$ &
$\omega^{P,V}_{\frac{1}{2}\frac{1}{2}}$ &  $\omega^{P,V}_{20}$ &$\omega^{P,V}_{02}$ &  $\omega^{P,V}_{11}$ &
$\omega^{P,V}_{\frac{3}{2}\frac{1}{2}}$   &  $\omega^{P,V}_{\frac{1}{2}\frac{3}{2}}$   \\
meson  &     &   GeV  &   GeV$^{3/2}$  &    &     &    &    &      &     &      &    \\    \hline
$B_{\bar bc}$&$1^1S_0$ & 6.276 & 0.250 & -0.0489 & -0.0060  & 0.0171  & 0.0049  & 0.0001  & 0.0006   & 0.0017    &   0.0002     \\  \hline
$B^\ast_{\bar bc}$  & $1^3S_1$  & 6.317 & 0.211 & -0.0540  & -0.0066   &  0.0188  & 0.0053  &  0.0001   & 0.0007    &  0.0019 & 0.0002  \\  \hline
\end{tabular}
\end{ruledtabular}
\end{table}

\begin{table}[h]
\caption{The comparison of obtained results for the production cross sections with nonrelativistic calculation.
In third column we present nonrelativistic result obtained in our model.}
\bigskip
\label{tb2}
\begin{ruledtabular}
\begin{tabular}{|c|c|c|c|}
Final state & Center-of-mass&Nonrelativistic  cross   & Relativistic cross \\
$B_{\bar bc}B_{b\bar c}$               &  energy s            &   section    $\sigma_{nr} $     &  section    $\sigma_r $ \\  \hline
$B^+_{\bar bc}+B^-_{b\bar c}$ & 22.0 GeV & 0.10 fb    &  0.03 fb  \\  \hline
$B^{\ast +}_{\bar bc}+B^{-}_{b\bar c}$ & 22.0 GeV& 0.10 fb     & 0.04 fb\\  \hline
$B^{\ast +}_{\bar bc}+B^{\ast -}_{b\bar c}$ & 22.0 GeV& 2.14 fb     &  0.58  fb \\  \hline
\end{tabular}
\end{ruledtabular}
\end{table}

\begin{figure}[t!]
\centering
\includegraphics[width=7.5 cm]{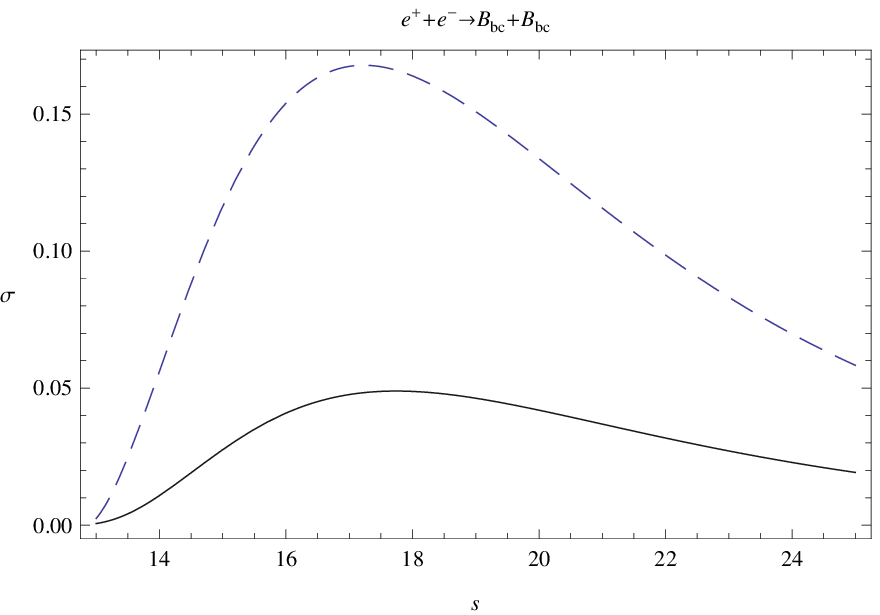}
\includegraphics[width=7.5 cm]{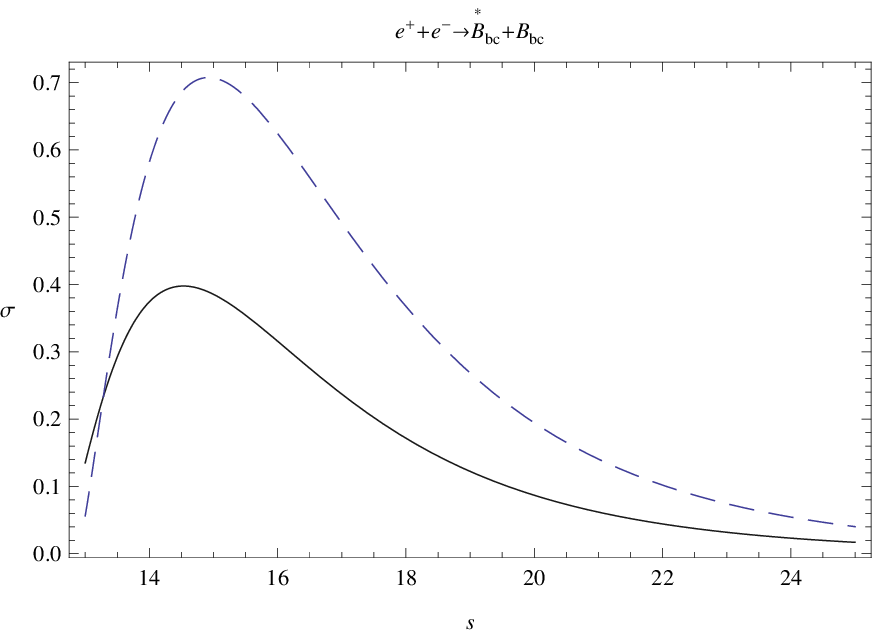}
\includegraphics[width=7.5 cm]{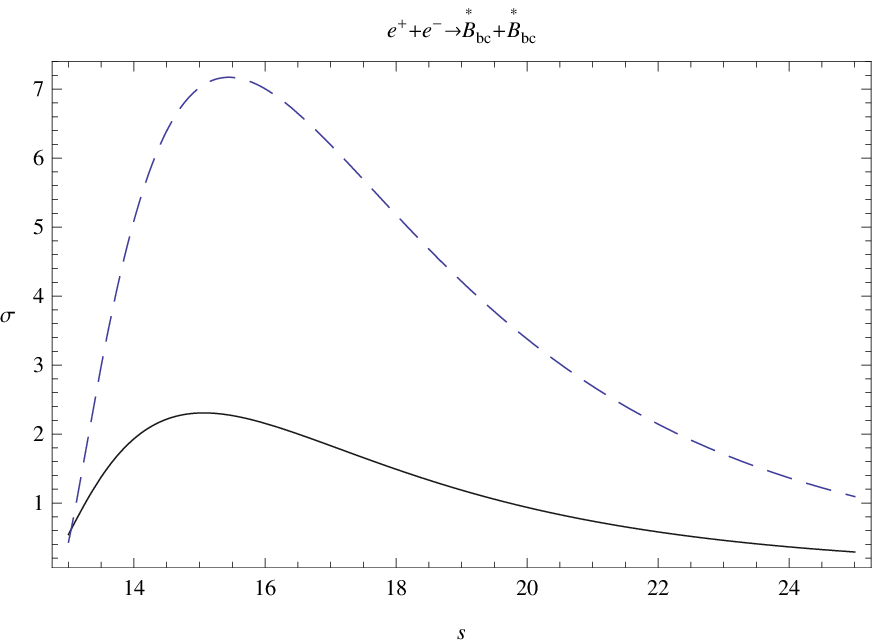}
\caption{The cross section in fb of $e^+e^-$ annihilation into a pair
of pseudoscalar and vector $B_c$ meson states as a function of the center-of-mass energy
$s$ (solid line). The dashed line shows nonrelativistic result without
bound state and relativistic corrections.}
\label{fig:fig2}
\end{figure}

To calculate relativistic corrections to the pseudoscalar and vector $B_c$-meson wave functions
$\Psi^0_{B_{\bar bc}}({\bf p})$ we use the Breit potential~\eqref{eq:breit} and
construct the effective potential model as in~\cite{EM2010,Lucha} by means of
the rationalization of kinetic energy operator.
The numerical values of the relativistic parameters entering the cross sections~\eqref{eq:sech1}, \eqref{eq:sech2} and \eqref{eq:sech3}
are obtained by the numerical solution of the Schr\"odinger equation \cite{LS}.
They are collected in Table~\ref{tb1}. We tested this model, calculating the masses of charmonium, bottomonium
and $B_c$ mesons and compare the results with existing experimental data and other theoretical predictions.
We can say that our results are in good agreement with them (the accuracy amounts to about one percent).
For example, in the case of low lying $(b\bar c)$ mesons
we obtain $M(1^1S_0)=6.276$ GeV and $M(1^3S_1)=6.317$ GeV.
Experimentally observed parameters of $B_c$ meson are consistent with different theoretical predictions \cite{PDG,rqm1,godfrey,glko}.
Numerical data related with charmonium states are discussed in~\cite{EM2010}.  One could argue that our estimates
of the charmonium mass spectrum agree with experimental data with more than a per cent accuracy \cite{EM2010,PDG}.
Our nonrelativistic values of $B_c$ meson wave functions at the origin differs on 20 per cent from the values presented
in \cite{rqm1,godfrey,glko} because we take $\alpha_s$=0.265 in this approximation.
Then we calculate the parameters of $B_c$ mesons and production cross sections as functions of
center-of-mass energy $s$.
Total cross section plots for the pair production of $B_c$ mesons are presented in Fig.~\ref{fig:fig2}.
In Table~\ref{tb2} we give numerical values of total production cross sections at certain center-of-mass
energies $s$ and compare them with nonrelativistic result in our quark model.
The effect of relativistic corrections to the bound state wave functions (the Breit potential)
plays a key role in total decreasing of the production cross sections as compared with nonrelativistic
results. The decreasing factor in the cross sections when passing from
nonrelativistic to relativistic results is equal approximately 3.
Total numerical results could be considered as an estimate for the experimental search.

\section{Numerical results and discussion}

Previous researches of the pair quarkonium production convincingly show that the calculation of the observed cross
sections without account of relativistic effects gives wrong results. Therefore, in our study, we focus on the inclusion
of relativistic corrections to the production cross sections within the previously developed formalism in the quark
model. The important role of relativistic effects in exclusive processes of pair production in electron-positron
annihilation is confirmed in the case of the $B_c$ mesons. Our construction of the amplitude of this process \eqref{eq:amp}
is aimed at keeping the two types of relativistic corrections. The corrections of the first type can be called the
relativistic corrections to the production amplitude connected with the relative quark momenta  ${\bf p}$ and ${\bf q}$.
The corrections of second type appear from the perturbative and nonperturbative
treatment of the quark-quark interaction operator which leads to the modification of the quark bound state wave
functions $\Psi^0_{B_{\bar bc}}({\bf p})$ as compared with nonrelativistic case.
We also systematically account for the bound state corrections working with masses of $B_c$ mesons.
The calculated masses of $B_c$ mesons agree well with previous theoretical results and experimental data
\cite{PDG,rqm1,godfrey,glko}. Note that the quark model, which we have used in the calculations is based
on quantum chromodynamics and has certain characteristics of universality.

Total cross sections for the exclusive pair production of pseudoscalar and vector $B_c$ mesons
in $e^+e^-$ annihilation are obtained from \eqref{eq:sech1}, \eqref{eq:sech2} and \eqref{eq:sech3}
after angular integration in the form:
\begin{equation}
\label{eq:sech1t}
\sigma_{PP}=\frac{4096\pi^3\alpha^2}{27s^{10}}M_P^2|\Psi^0_{B_{\bar bc}}(0)|^2|\Psi_{{B}_{b\bar c}}(0)|^2
\bigl(1-\frac{4M^2_{B_{\bar bc}}}{s^2}\bigr)^{3/2}\left[\frac{Q_c\alpha_s(\frac{m_2^2}{M^2}s^2)}{r_2^3}F_{1}^P-
\frac{Q_b\alpha_s(\frac{m_1^2}{M^2}s^2)}{r_1^3}F_{2}^P
\right]^2,
\end{equation}
\begin{equation}
\label{eq:sech2t}
\sigma_{PV}=\frac{2048\pi^3\alpha^2}{27s^8}\frac{M_PM_V}{M^2}
\left[\left(1-\frac{(M_P+M_V)^2}{s^2}\right)\left(1-\frac{(M_P-M_V)^2}{s^2}\right)\right]^{3/2}\times
\end{equation}
\begin{displaymath}
|\Psi^0_{B^\ast_{\bar bc}}(0)|^2|\Psi_{{B}_{b\bar c}}(0)|^2
\left[\frac{Q_c\alpha_s\left(\frac{m_2^2}{M^2}s^2\right)}{r_2^3}F_{1}^{P,V}+
\frac{Q_b\alpha_s\left(\frac{m_1^2}{M^2}s^2\right)}{r_1^3}F_{2}^{P,V}\right]^2,
\end{displaymath}
\begin{equation}
\label{eq:sech3t}
\sigma_{VV}=\frac{512\pi^3\alpha^2}{27s^{10}}M_V^2|\Psi^0_{B^\ast_{\bar bc}}(0)|^2|\Psi_{B^\ast_{b\bar c}}(0)|^2
\left(1-\frac{4M^2_V}{s^2}\right)^{3/2}\left(3F_A-F_B\right).
\end{equation}

The plots of total cross sections for the production of
pseudoscalar+pseudoscalar, pseudoscalar+vector and vector+vector $B_c$ mesons
as functions of center-of-mass energy $s$ are presented in Fig.~\ref{fig:fig2}.
At present there are no experimental data about such exclusive processes.
Therefore, these graphs can serve in the future as a guide in conducting relevant experiments.
Comparing three plots in Fig.~\ref{fig:fig2}, we see that the greatest values of production cross
sections correspond to the pair vector $B_c$ meson production.
Assuming that a luminosity at the B-factory ${\mathcal L}=10^{34}~cm^{-2}\cdot c^{-1}$ the yield of pairs of
vector $B_c$ mesons can be near $50$ events per month at the center-of-mass energy $s=16~GeV$.
As it follows from Fig.~\ref{fig:fig2} an account of relativistic and bound state corrections decreases the cross
section value as compared with nonrelativistic result. It is important to emphasize that we call
nonrelativistic result that one which is obtained with pure nonrelativistic Hamiltonian from \eqref{eq:breit}
and the bound state mass $M_{B_{\bar bc}}=m_1+m_2$.
There are several important factors that significantly affect the total value of cross-sections
in the transition from non-relativistic description to a relativistic theory although they act in
different directions.
Relativistic corrections to the production amplitude increase nonrelativistic result on a few tenths of per cents.
But another relativistic corrections to the bound state wave functions and bound state corrections have an opposite effect.
The value of bound state wave function at the origin is significantly reduced when taking into account relativistic
corrections and, as a result, reduced the production cross sections in the case of pair P+P, P+V and V+V $B_c$ mesons.

We presented a treatment of relativistic effects in the production of $B_c$ meson pair in $e^+e^-$ annihilation.
Our approach to the study of these processes is based on the quark model, which is a microscopic theory of
quark-gluon interaction. In contrast to NRQCD formalism \cite{BBL} the quark model allows you to perform the calculation
of various parameters describing the formation of bound states of heavy quarks.
In the days of the dominant concept of total parameterization of strong interaction processes, which,
incidentally, does not allow to find these parameters with a high degree of accuracy, this property
of the quark model looks like its advantage.
We distinguish two types of relativistic contributions to the production amplitudes \eqref{eq:amp11},
\eqref{eq:amp22}, \eqref{eq:amp33}.
The first type consists of relativistic $p/m_{1,2}$, $q/m_{1,2}$ corrections to the wave functions and their
relativistic transformations. The second type includes relativistic $p/s$, $q/s$ corrections arising from the
expansion of the quark and gluon propagators. The latter corrections are taken into account up
to the second order. Note that the expansion parameter $p/s$ is very small. In our analysis
of the production amplitudes we correctly take into account
relativistic contributions of order $O(p^2/m_{1,2}^2)$ for the $B_c$ mesons. Therefore the first basic
theoretical uncertainty of our calculation is connected with omitted terms of order $O({p}^4/m_{1,2}^4)$.
An estimate of the possible magnitude of these corrections can be given using the average value
of heavy quark velocity squared in the charmonium $\langle v^2\rangle=0.3$,
We expect that relativistic corrections
of order $O({p}^4/m_{1,2}^4)$ to the cross sections~\eqref{eq:sech1t}, \eqref{eq:sech2t}, \eqref{eq:sech3t},
coming from the production amplitude should not exceed $30\%$ of the obtained relativistic result.
Another important theoretical uncertainty is related with the bound state wave functions at
relativistic momenta of order of quark masses. We consider that total error of the wave function
determination in this region amounts to $5\%$. Then the corresponding error in the cross sections~~\eqref{eq:sech1t},
\eqref{eq:sech2t}, \eqref{eq:sech3t} is not exceeding $20\%$. Of course, this estimate
is very approximate one but it agrees with the calculation of quarkonium masses with the accuracy
better than one per cent.
Finally, an important part of total theoretical error is connected with
radiative corrections of order $\alpha_s$ which are omitted in our work.
In our calculation of the cross sections we account for effectively only some part
of one loop corrections by means of the Breit Hamiltonian.
Therefore, we assume that the radiative corrections of order $O(\alpha_s)$ can produce $20\%$ change of the production cross sections.
Our total maximum theoretical errors are estimated in $40\%$.
To obtain this estimate we add the above mentioned uncertainties in quadrature.

\acknowledgments
The authors are grateful to D.~Ebert, R.N.~Faustov and V.O.~Galkin for useful discussions.
The work is supported by the Ministry of Education and
Science of Russia under Competitiveness Enhancement Program 2016-2020 and grant No. 1394.

\appendix

\section{The coefficient functions $F_{i}^{\cal P}$, $F_{i}^{\cal P,V}$ and $F_{i}^{\cal V}$ entering in
the $B_c$ meson production amplitudes (13)-(15)}

General structure of the $B_c$ meson  production amplitudes studied in
this work is the following:
\bga
\label{eq:A1}
\mathcal M=\frac{8\pi^2\alpha}{3s}\sqrt{M_{B_{\bar bc}}M_{B_{b\bar c}}}\,[\bar v(p_+)
\gamma_\beta u(p_-)]\times \\
\int\!\frac{d\mathbf{p}}{(2\pi)^3}\int\!\frac{d\mathbf{q}}{(2\pi)^3}
\frac{\Psi^0_{B_{\bar bc}}({\bf p})}
{\sqrt{\frac{\epsilon_1(p)}{m_1}\frac{(\epsilon_1(p)+m_1)}{2m_1}\frac{\epsilon_2(p)}{m_2}
\frac{(\epsilon_2(p)+m_2)}{2m_2}}}
\frac{\Psi^0_{B_{b\bar c}}({\bf q})}
{\sqrt{\frac{\epsilon_1(q)}{m_1}\frac{(\epsilon_1(q)+m_1)}{2m_1}\frac{\epsilon_2(q)}{m_2}
\frac{(\epsilon_2(q)+m_2)}{2m_2}}}\times \\
\mathrm{Tr}\bigl\{\mathcal T_{12}^\beta+\mathcal T_{34}^\beta\bigr\},
\ega

\bga
\mathcal T_{12}^\beta=\mathcal Q_c\alpha_b\Bigl[\frac{\hat v_1-1}2+\hat v_1\frac{\mathbf p^2}
{2m_2(\epsilon_2(p)+m_2)}-\frac{\hat p}{2m_2}\Bigr]\Sigma^{(1)}_{P,V}(1+\hat v_1)\times\\
\Bigl[\frac{\hat v_1+1}2+\hat v_1\frac{\mathbf p^2}{2m_1(\epsilon_1(p)+m_1)}+\frac{\hat p}{2m_1}
\Bigr]\left[\gamma^\beta\frac{\hat p_1-\hat l+m_1}{(l-p_1)^2-m_1^2}\,\gamma_\mu+\gamma_\mu\,\frac{\hat l-\hat q_1+m_1}{(l-q_1)^2-m_1^2}\gamma^\beta\right]D^{\mu\nu}(k_2)\times \\
\Bigl[\frac{\hat v_2-1}2+\hat v_2\frac{\mathbf q^2}{2m_1(\epsilon_1(q)+m_1)}+\frac{\hat q}{2m_1}
\Bigr]\Sigma^{(2)}_{P,V}(1+\hat v_2)
\Bigl[\frac{\hat v_2+1}2+\hat v_2\frac{\mathbf q^2}{2m_2(\epsilon_2(q)+m_2)}-\frac{\hat q}{2m_2}\Bigr]\gamma_\nu,
\ega

\bga
\mathcal T_{34}^\beta=\mathcal Q_b\alpha_c\Bigl[\frac{\hat v_1-1}2+\hat v_1
\frac{\mathbf p^2}{2m_1(\epsilon_1(p)+m_1)}+\frac{\hat p}{2m_1}\Bigr]\Sigma^{(1)}_{P,V}(1+\hat v_1)\times\\
\Bigl[\frac{\hat v_1+1}2+\hat v_1\frac{\mathbf p^2}{2m_2(\epsilon_2(p)+m_2)}-\frac{\hat p}{2m_2}\Bigr]
\left[
\gamma^\beta\frac{\hat p_2-\hat l+m_2}{(l-p_2)^2-m_2^2}\,\gamma_\mu+
\gamma_\mu\frac{\hat l-\hat q_2+m_2}{(l-q_2)^2-m_2^2}\,\gamma^\beta
\right]
D^{\mu\nu}(k_1)\times \\
\Bigl[\frac{\hat v_2-1}2+\hat v_2\frac{\mathbf q^2}{2m_2(\epsilon_2(q)+m_2)}-
\frac{\hat q}{2m_2}\Bigr]\Sigma^{(2)}_{P,V}(1+\hat v_2)
\Bigl[\frac{\hat v_2+1}2+\hat v_2\frac{\mathbf q^2}{2m_1(\epsilon_1(q)+m_1)}+
\frac{\hat q}{2m_1}\Bigr]\gamma_\nu,
\ega
where $\Sigma^{(1),(2)}_{P,V}$ is equal to $\gamma_5$ for pseudoscalar $B_c$ meson and
$\hat\varepsilon_{\mathcal{V}}$ for vector $B_c$ meson. The trace calculation in \eqref{eq:A1}
leads to amplitudes ${\cal M}_{PP}$, ${\cal M}_{PV}$ and ${\cal M}_{VV}$ presented in~\eqref{eq:amp11}-\eqref{eq:amp33}.
Corresponding functions $F_{i}^{\cal P}$, $F_{i}^{\cal P,V}$ and $F_{1i}^{\cal V}$ are written below in the used approximation.

\vspace{5mm}

{\underline {$e^++e^-\to B_{\bar bc}^++B_{b\bar c}^{-}$}.

\begin{equation}
\label{eq:aa1}
F_{1}^{\cal P}=s^{-2}\omega_{\frac{3}{2}\frac{1}{2}}(- 8r_1^2r_2^{-1} - 4r_1^2)
+ s^{-2}\omega_{\frac{1}{2}\frac{3}{2}} (  - 8r_1^2r_2^{-1} - 4r_1^2 )
+ s^{-2}\omega_{\frac{1}{2}\frac{1}{2}} (  - 8r_1^2r_2^{-1} - 4r_1^2 )+
\end{equation}
\begin{displaymath}
s^{-2}\omega^2_{\frac{1}{2}\frac{1}{2}} (  - \frac{4}{3}r_1 - \frac{88}{3}r_1^2r_2^{-1} - 12r_1^2 +
\frac{8}{3}r_1^3r_2^{-2} )
+ \omega_{\frac{3}{2}\frac{1}{2}} ( \frac{4}{3} - 2r_2 + \frac{26}{3}r_1r_2^{-1} - \frac{17}{3}r_1 - 8r_1^2
r_2^{-1} - 2r_1^2 )+
\end{displaymath}
\begin{displaymath}
2r_1(\omega_{11}+ \omega_{02} + \omega_{20}) + r_1(\omega_{01}+ \omega_{10})^2
 + \omega_{\frac{1}{2}\frac{3}{2}}( \frac{2}{3} - r_2 + \frac{40}{3}r_1r_2^{-1} - \frac{26}{3}r_1 - 12r_1^2
 r_2^{-1} - 2r_1^2 )+
\end{displaymath}
\begin{displaymath}
\omega_{\frac{1}{2}\frac{1}{2}}^2 ( \frac{46}{9} - \frac{17}{9}r_2 - \frac{38}{9}r_1r_2^{-1} - \frac{59}{3}r_1 + \frac{4}{9}
 r_1^2r_2^{-2} + \frac{112}{9}r_1^2r_2^{-1} + 2r_1^2 - \frac{8}{9}r_1^3r_2^{-2})+
\end{displaymath}
\begin{displaymath}
\omega_{\frac{1}{2}\frac{1}{2}} ( \frac{2}{3} - r_2 + \frac{26}{3}r_1r_2^{-1} - \frac{8}{3}r_1 - 8r_1^2
r_2^{-1} - 2r_1^2 )+ s^2\omega_{\frac{3}{2}\frac{1}{2}} (  - \frac{2}{3}r_2 - r_1r_2^{-1} - r_1 + r_1^2r_2^{-1} )
- s^2r_2(\omega_{11}+\omega_{02}+\omega_{20})-
 \end{displaymath}
\begin{displaymath}
\frac{1}{2}r_2s^2(\omega_{01}+\omega_{10})^2+ s^2\omega_{\frac{1}{2}\frac{3}{2}}
(- 1 - \frac{1}{3}r_2 - 2r_1r_2^{-1} + \frac{1}{3}r_1 + 2r_1^2r_2^{-1} )+ s^2\omega_{\frac{1}{2}\frac{1}{2}}
(- \frac{1}{3}r_2 - r_1r_2^{-1} - r_1 + r_1^2r_2^{-1} )+
\end{displaymath}
\begin{displaymath}
s^2\omega_{\frac{1}{2}\frac{1}{2}}^2 (\frac{1}{9}r_2- \frac{13}{18} + \frac{13}{9}r_1r_2^{-1} + \frac{47}{18}r_1
-\frac{2}{9}r_1^2r_2^{-2} - \frac{19}{9}r_1^2r_2^{-1} + \frac{2}{9}r_1^3r_2^{-2} )
+Z_1Y_1(-\frac{M_p}{2M}\eta_1-\frac{m_1}{2M}+s^2\frac{M}{2M_P}\eta_2).
\end{displaymath}

\vspace{5mm}

{\underline {$e^++e^-\to B^{\ast +}_{\bar bc}+B^-_{b\bar c}$}.
\begin{equation}
\label{eq:aa2}
F_{1}^{\cal P,V}=Y_1\left(\frac{1}{2}Z_2\frac{M_V}{M}-\frac{1}{2}Z_1\frac{M_V}{M}\eta_2+Z_1\frac{M_P}{M}-
\frac{1}{2}\frac{m_1}{M}Z_1\right)
+ s^{-2}\omega^V_{\frac{3}{2}\frac{1}{2}}  ( 4r_1r_2^{-1} - \frac{2}{3}r_1 )+
\end{equation}
\begin{displaymath}
s^{-2}\omega^V_{\frac{1}{2}\frac{3}{2}}  ( 4r_1r_2^{-1} + 2r_1 )
+ s^{-2}\omega^V_{\frac{1}{2}\frac{1}{2}}  ( 4r_1r_2^{-1} + 2r_1 )
+ s^{-2}\omega^P_{\frac{3}{2}\frac{1}{2}} ( 4r_1r_2^{-1} - \frac{2}{3}r_1 )
+ s^{-2}\omega^P_{\frac{1}{2}\frac{3}{2}}  ( 4r_1r_2^{-1} + 2r_1 )+
\end{displaymath}
\begin{displaymath}
s^{-2}\omega^P_{\frac{1}{2}\frac{1}{2}}  ( 4r_1r_2^{-1} + 2r_1 )
+ s^{-2}\omega^P_{\frac{1}{2}\frac{1}{2}}\omega^V_{\frac{1}{2}\frac{1}{2}} ( \frac{8}{9} + \frac{8}{9}r_2 + \frac{8}{9}r_1r_2^{-1} + \frac{32}{9}r_1 - \frac{8}{3}r_1^2r_2^{-2} + \frac{8}{3}r_1^2r_2^{-1} )+
\end{displaymath}
\begin{displaymath}
\omega^V_{\frac{3}{2}\frac{1}{2}} ( 4 - \frac{2}{3}r_2 - \frac{1}{3}r_1r_2^{-1} - \frac{1}{3}r_1 )
- \omega^V_{11}- \omega^V_{02}- \omega^V_{20}
+ \omega^V_{\frac{1}{2}\frac{3}{2}} ( \frac{5}{3} - \frac{1}{3}r_2 - \frac{2}{3}r_1r_2^{-1} + \frac{4}{3}r_1 )+
\end{displaymath}
\begin{displaymath}
\omega^V_{\frac{1}{2}\frac{1}{2}} ( 2 - \frac{1}{3}r_2 - \frac{1}{3}r_1r_2^{-1} + \frac{2}{3}r_1 )
+\omega^P_{\frac{3}{2}\frac{1}{2}}( \frac{8}{3} - \frac{1}{3}r_1r_2^{-1} )
-\omega^P_{11}- \omega^P_{02}- \omega^P_{20}
-\omega^P_{01}\omega^V_{10}- \omega^P_{01}\omega^V_{01}- \omega^P_{10}\omega^V_{10}-
\end{displaymath}
\begin{displaymath}
\omega^P_{10}\omega^V_{01}
+\omega^P_{\frac{1}{2}\frac{3}{2}} ( \frac{7}{3} - \frac{2}{3}r_1r_2^{-1} )
+\omega^P_{\frac{1}{2}\frac{1}{2}} ( \frac{4}{3} - \frac{1}{3}r_1r_2^{-1} )
+\omega^P_{\frac{1}{2}\frac{1}{2}}\omega^V_{\frac{1}{2}\frac{1}{2}}
(-\frac{1}{3} + \frac{1}{9}r_2-\frac{22}{9}r_1r_2^{-1} + \frac{4}{9}r_1^2r_2^{-2}-\frac{4}{9}r_1^2r_2^{-1} ).
\end{displaymath}

\vspace{5mm}

{\underline {$e^++e^-\to B^{\ast +}_{\bar bc}+B^{\ast -}_{b\bar c}$}.

\begin{equation}
F_{1}^{\cal V,\beta}=F_{11}^{\cal V}(v_1^\beta-v_2^\beta)(\epsilon_1\epsilon_2)+
F_{12}^{\cal V}(v_1^\beta-v_2^\beta)(\epsilon_1v_2)(\epsilon_2v_1)+
F_{13}^{\cal V}[\varepsilon_1^\beta(\varepsilon_2 v_1)-\varepsilon_2^\beta(\varepsilon_1 v_2)],
\end{equation}
\begin{displaymath}
F_{11}^{\cal V}=\frac{1}{2}\left(\frac{m_1}{M}+\frac{M_V}{M}\eta_1\right)Z_1Y_1
+\frac{4}{3}s^{-2}r_1\omega_{\frac{1}{2}\frac{1}{2}}^2 + 8s^{-2}r_1^2r_2^{-1}\omega_{\frac{3}{2}\frac{1}{2}} +
8s^{-2}r_1^2r_2^{-1}\omega_{\frac{1}{2}\frac{3}{2}}+8s^{-2}r_1^2r_2^{-1}\omega_{\frac{1}{2}\frac{1}{2}}-
\end{displaymath}
\begin{displaymath}
\frac{56}{9}s^{-2}r_1^2r_2^{-1}\omega_{\frac{1}{2}\frac{1}{2}}^2 +
4s^{-2}r_1^2\omega_{\frac{3}{2}\frac{1}{2}} + 4s^{-2}r_1^2\omega_{\frac{1}{2}\frac{3}{2}} +
4s^{-2}r_1^2\omega_{\frac{1}{2}\frac{1}{2}} + \frac{4}{3}s^{-2}r_1^2\omega_{\frac{1}{2}\frac{1}{2}}^2 -
\frac{8}{3}s^{-2}r_1^3r_2^{-2}\omega_{\frac{1}{2}\frac{1}{2}}^2 -
\end{displaymath}
\begin{displaymath}
\frac{4}{3}\omega_{\frac{3}{2}\frac{1}{2}} + \frac{2}{3}\omega_{\frac{1}{2}\frac{3}{2}} -
\frac{2}{3}\omega_{\frac{1}{2}\frac{1}{2}} - \frac{2}{9}\omega_{\frac{1}{2}\frac{1}{2}}^2 + 2r_2\omega_{\frac{3}{2}\frac{1}{2}} +
r_2\omega_{\frac{1}{2}\frac{3}{2}} +
r_2\omega_{\frac{1}{2}\frac{1}{2}} +
\frac{5}{9}r_2\omega_{\frac{1}{2}\frac{1}{2}}^2 - \frac{2}{3}r_1r_2^{-1}\omega_{\frac{3}{2}\frac{1}{2}} -
\frac{4}{3}r_1r_2^{-1}\omega_{\frac{1}{2}\frac{3}{2}} -
\end{displaymath}
\begin{displaymath}
\frac{2}{3}r_1r_2^{-1}\omega_{\frac{1}{2}\frac{1}{2}} - 2r_1r_2^{-1}\omega_{\frac{1}{2}\frac{1}{2}}^2 +
5r_1\omega_{\frac{3}{2}\frac{1}{2}} - 2r_1\omega_{11} -
2r_1\omega_{02} - 2r_1\omega_{20} - r_1\omega_{01}^2 - 2r_1\omega_{10}\omega_{01} - r_1\omega_{10}^2 +
4r_1\omega_{\frac{1}{2}\frac{3}{2}} +
\end{displaymath}
\begin{displaymath}
\frac{10}{3}r_1\omega_{\frac{1}{2}\frac{1}{2}} + \frac{11}{9}r_1\omega_{\frac{1}{2}\frac{1}{2}}^2 +
\frac{20}{9}r_1^2r_2^{-1}\omega_{\frac{1}{2}\frac{1}{2}}^2 + \frac{4}{9}r_1^3r_2^{-2}\omega_{\frac{1}{2}\frac{1}{2}}^2 ),
\end{displaymath}

\begin{displaymath}
F_{12}^{\cal V}=-\frac{8}{9}s^{-2}\omega_{\frac{1}{2}\frac{1}{2}}^2 +\frac{8}{9}s^{-2}r_2\omega_{\frac{1}{2}\frac{1}{2}}^2 +
4s^{-2}r_1r_2^{-1}\omega_{\frac{3}{2}\frac{1}{2}} + 8s^{-2}r_1r_2^{-1}
\omega_{\frac{1}{2}\frac{3}{2}} +4s^{-2}r_1r_2^{-1}\omega_{\frac{1}{2}\frac{1}{2}} +
\end{displaymath}
\begin{displaymath}
\frac{20}{9}s^{-2}r_1r_2^{-1}\omega_{\frac{1}{2}\frac{1}{2}}^2 + 8s^{-2}r_1\omega_{\frac{3}{2}\frac{1}{2}} +
4s^{-2}r_1\omega_{\frac{1}{2}\frac{3}{2}} + 4s^{-2}
r_1\omega_{\frac{1}{2}\frac{1}{2}} - \frac{28}{9}s^{-2}r_1\omega_{\frac{1}{2}\frac{1}{2}}^2 +
\frac{16}{9}s^{-2}r_1^2r_2^{-2}\omega_{\frac{1}{2}\frac{1}{2}}^2 -
\end{displaymath}
\begin{displaymath}
4s^{-2}r_1^2r_2^{-1}\omega_{\frac{3}{2}\frac{1}{2}} -8s^{-2}r_1^2
r_2^{-1}\omega_{\frac{1}{2}\frac{3}{2}} - 4s^{-2}r_1^2r_2^{-1}\omega_{\frac{1}{2}\frac{1}{2}} -
\frac{20}{9}s^{-2}r_1^2r_2^{-1}\omega_{\frac{1}{2}\frac{1}{2}}^2 -
\frac{16}{9}s^{-2}r_1^3r_2^{-2}\omega_{\frac{1}{2}\frac{1}{2}}^2,
\end{displaymath}
\begin{displaymath}
F_{13}^{\cal V}=\left(\frac{M_V}{M}-\frac{M_V}{2M}\eta_1+\frac{m_1}{2M}\right)Z_1Y_1 + \frac{8}{9}s^{-2}\omega_{\frac{1}{2}\frac{1}{2}}^2 +
\frac{8}{9}s^{-2}r_2\omega_{\frac{1}{2}\frac{1}{2}}^2 + 8s^{-2}r_1r_2^{-1}\omega_{\frac{3}{2}\frac{1}{2}} +
8s^{-2}r_1r_2^{-1}\omega_{\frac{1}{2}\frac{3}{2}} +
\end{displaymath}
\begin{displaymath}
8s^{-2}r_1r_2^{-1}\omega_{\frac{1}{2}\frac{1}{2}} +
\frac{8}{9}s^{-2}r_1r_2^{-1}\omega_{\frac{1}{2}\frac{1}{2}}^2 - \frac{4}{3}s^{-2}r_1\omega_{\frac{3}{2}\frac{1}{2}} +
4s^{-2}r_1\omega_{\frac{1}{2}\frac{3}{2}} + 4s^{-2}r_1\omega_{\frac{1}{2}\frac{1}{2}} + \frac{8}{3}s^{-2}r_1\omega_{\frac{1}{2}\frac{1}{2}}^2 -
\end{displaymath}
\begin{displaymath}
\frac{8}{3}s^{-2}r_1^2r_2^{-2}\omega_{\frac{1}{2}\frac{1}{2}}^2 -
\frac{8}{9}s^{-2}r_1^2r_2^{-1}\omega_{\frac{1}{2}\frac{1}{2}}^2 + 4\omega_{\frac{3}{2}\frac{1}{2}} - 2\omega_{11} - 2\omega_{02} -
2\omega_{20} - (\omega_{01}+\omega_{10})^2 + \frac{8}{3}\omega_{\frac{1}{2}\frac{3}{2}} + 2\omega_{\frac{1}{2}\frac{1}{2}} -
\end{displaymath}
\begin{displaymath}
\frac{1}{3}\omega_{\frac{1}{2}\frac{1}{2}}^2 + \frac{2}{3}r_2\omega_{\frac{3}{2}\frac{1}{2}} + \frac{1}{3}r_2\omega_{\frac{1}{2}\frac{3}{2}} + \frac{1}{3}r_2\omega_{\frac{1}{2}\frac{1}{2}} +
\frac{1}{3}r_2\omega_{\frac{1}{2}\frac{1}{2}}^2 - \frac{2}{3}r_1r_2^{-1}\omega_{\frac{3}{2}\frac{1}{2}} - \frac{4}{3}r_1r_2^{-1}\omega_{\frac{1}{2}\frac{3}{2}} -
\frac{2}{3}r_1r_2^{-1}\omega_{\frac{1}{2}\frac{1}{2}} -
\end{displaymath}
\begin{displaymath}
\frac{2}{3}r_1r_2^{-1}\omega_{\frac{1}{2}\frac{1}{2}}^2 + \frac{5}{3}r_1\omega_{\frac{3}{2}\frac{1}{2}} +
\frac{4}{3}r_1\omega_{\frac{1}{2}\frac{3}{2}} + \frac{2}{3}r_1\omega_{\frac{1}{2}\frac{1}{2}} + \frac{4}{9}r_1\omega_{\frac{1}{2}\frac{1}{2}}^2 +
\frac{4}{9}r_1^2r_2^{-2}\omega_{\frac{1}{2}\frac{1}{2}}^2 - \frac{4}{9}r_1^2r_2^{-1}\omega_{\frac{1}{2}\frac{1}{2}}^2.
\end{displaymath}

The functions $F_{2}^{\cal P,V}$ entering in the production amplitudes (13)-(15)} can be obtained from $F_{1}^{\cal P,V}$
changing $r_2\leftrightarrow r_1$, $m_1\leftrightarrow m_2$ and $\omega_{ij}\to\omega_{ji}$, $\eta_{1,2}\leftrightarrow \rho_{2,1}$,
$M_V\leftrightarrow M_P$.

\end{document}